\documentclass[11pt,a4paper]{article}

\usepackage[numbers, square, comma, sort&compress]{natbib}
\usepackage{amssymb}
\usepackage{subcaption}
\usepackage{amsthm}
\usepackage{amstext}
\usepackage{amsmath}
\usepackage{amsfonts}
\usepackage{empheq}
\usepackage{verbatim}
\usepackage{geometry}
\usepackage{latexsym}
\usepackage{t1enc}
\usepackage{graphicx}
\usepackage{color}
\usepackage{colortbl}
\usepackage[all]{xy}
\usepackage{makeidx}
\usepackage{slashed}
\usepackage{multicol}
\usepackage{alltt,color}
\usepackage[utf8]{inputenc}

\usepackage{hyperref}
\hypersetup{colorlinks=true, citecolor=blue, urlcolor=blue, linkcolor=blue}

\allowdisplaybreaks

\newcommand{\be}{\begin{equation}}
\newcommand{\ee}{\end{equation}}
\newcommand{\bea}{\begin{eqnarray}}
\newcommand{\eea}{\end{eqnarray}}

\def\beq{\begin{equation}}
\def\eeq{\end{equation}}
\def\beqa{\begin{eqnarray}}
\def\eeqa{\end{eqnarray}}

\def\spa#1.#2{\left\langle#1\,#2\right\rangle}
\def\spb#1.#2{\left[#1\,#2\right]}
\def\spash#1.#2{\spa{\smash{#1}}.{\smash{#2}}}
\def\spbsh#1.#2{\spb{\smash{#1}}.{\smash{#2}}}
\def\sand#1.#2.#3{%
  \left\langle\smash{#1^{-}}{\vphantom1}\right|{#2}%
  \left|\smash{#3^{-}}{\vphantom1}\right\rangle}
\def\sandp#1.#2.#3{%
  \left\langle\smash{#1^{-}}{\vphantom1}\right|{#2}%
  \left|\smash{#3^{+}}{\vphantom1}\right\rangle}
\def\sandpp#1.#2.#3{%
  \left\langle\smash{#1^{+}}{\vphantom1}\right|{#2}%
  \left|\smash{#3^{+}}{\vphantom1}\right\rangle}
\def\sandpm#1.#2.#3{%
  \left\langle\smash{#1^{+}}{\vphantom1}\right|{#2}%
  \left|\smash{#3^{-}}{\vphantom1}\right\rangle}
\def\sandmp#1.#2.#3{%
  \left\langle\smash{#1^{-}}{\vphantom1}\right|{#2}%
  \left|\smash{#3^{+}}{\vphantom1}\right\rangle}

\def\ssand#1.#2.#3{%
  \left\langle\smash{#1}{\vphantom1}\right|{#2}%
  \left|\smash{#3}{\vphantom1}\right]}
\def\ssandp#1.#2.#3{%
  \left\langle\smash{#1}{\vphantom1}\right|{#2}%
  \left|\smash{#3}{\vphantom1}\right\rangle}
\def\ssandpp#1.#2.#3{%
  \left\langle\smash{#1}{\vphantom1}\right|{#2}%
  \left|\smash{#3}{\vphantom1}\right\rangle}

\def\proj{\flat}
\def\projdot#1.#2{k_{#1}^\proj\cdot k_{#2}^\proj}
\def\sandff#1.#2.#3{%
  \left\langle\smash{#1^{\proj,-}}{\vphantom1}\right|{#2}%
  \left|\smash{#3^{\proj,-}}{\vphantom1}\right\rangle}
\def\sandnf#1.#2.#3{%
  \left\langle\smash{#1^{-}}{\vphantom1}\right|{#2}%
  \left|\smash{#3^{\proj,-}}{\vphantom1}\right\rangle}
\def\sandfn#1.#2.#3{%
  \left\langle\smash{#1^{\proj,-}}{\vphantom1}\right|{#2}%
  \left|\smash{#3^{-}}{\vphantom1}\right\rangle}

\def\spa#1.#2{\left\langle#1\,#2\right\rangle}
\def\spb#1.#2{\left[#1\,#2\right]}

\numberwithin{equation}{section}

\begin{document}

\begin{titlepage}

\hbox{QMUL-PH-18-18}

\vskip 25mm

\begin{center}
\Large{\sc{Effective Field Theory in the top sector: do multijets help?}}
\end{center}

\vskip 8mm

\begin{center}

Christoph Englert$^a$\footnote{Email:
    christoph.englert@glasgow.ac.uk}, Michael
  Russell$^b$\footnote{Email: russell@thphys.uni-heidelberg.de} ~and
Chris D. White$^c$\footnote{Email:
    christopher.white@qmul.ac.uk} \\ [6mm]

\vspace{6mm}

$^a$ \textit{SUPA, School of Physics and Astronomy, University of
    Glasgow,} \\ \textit{Glasgow, G12 8QQ, United Kingdom}\vspace{0.5cm} \\
$^b$  \textit{Institut f\"ur Theoretische Physik, Universit\"at Heidelberg, Germany}\vspace{0.5cm}\\
$^c$ \textit{Centre for Research in String Theory, School of Physics
    and Astronomy}, \\ \textit{Queen Mary University of London, 327 Mile End Road}, \\ \textit{London E1 4NS, United Kingdom} 

\end{center}

\vspace{5mm}

\begin{abstract}
\noindent
Many studies of possible new physics employ effective field theory
(EFT), whereby corrections to the Standard Model take the form of
higher-dimensional operators, suppressed by a large energy scale. Fits
of such a theory to data typically use parton level observables, which
limits the datasets one can use. In order to theoretically model
search channels involving many additional jets, it is important to
include tree-level matrix elements matched to a parton shower
algorithm, and a suitable matching procedure to remove the double
counting of additional radiation. There are then two potential
problems: (i) EFT corrections are absent in the shower, leading to an
extra source of discontinuities in the matching procedure; (ii) the
uncertainty in the matching procedure may be such that no additional
constraints are obtained from observables sensitive to radiation.  In
this paper, we review why the first of these is not a problem in
practice, and perform a detailed study of the second. In particular,
we quantify the additional constraints on EFT expected from top pair
plus multijet events, relative to inclusive top pair production alone.
\end{abstract}

\end{titlepage}

\section{Introduction}
\label{sec:introduction}

The search for physics beyond the Standard Model (BSM) is the most
pressing problem in particle physics, especially given the on-going
experimental programme at the Large Hadron Collider. To date, clear
signatures of BSM physics have remained elusive, although it is widely
suspected that the new physics may have something to do with the
nature of electroweak symmetry breaking, and thus affect the behaviour
of the recently discovered Higgs boson or, due to its large Yukawa
coupling with the former, the top quark. 

The lack of clear evidence for BSM physics thus far motivates the use
of effective field
theory~{\cite{Weinberg:1978kz,Buchmuller:1985jz,Burges:1983zg,Leung:1984ni,Hagiwara:1986vm,Grzadkowski:2010es}} (for a comprehensive review see \cite{Brivio:2017vri}),
in which one characterises corrections to the SM Lagrangian by
gauge-invariant higher dimensional operators built from the SM
fields. This has the advantage of being manifestly model independent,
but is only applicable if the lowest energy scale associated with the
new physics is above the typical energies probed by the collider of
interest. Given that this is likely to be a viable situation at the
LHC, such techniques have been widely used in the contexts of Higgs
and electroweak precision
physics~\cite{Pierce:2006dh,Alonso:2012pz,Djouadi:2012rh,Low:2012rj,Zhang:2012cd,Degrande:2013kka,Elias-Miro:2013gya,Contino:2013kra,Grojean:2013nya,Gripaios:2013lea,Hayreter:2013kba,Buchalla:2013eza,Alonso:2013hga,Gavela:2014uta,Wells:2014pga,Dawson:2014ora,Delgado:2014jda,Henning:2014wua,Ellis:2014jta,Gupta:2014rxa,Lehman:2014jma,Englert:2014ffa,Corbett:2015ksa,Chiang:2015ura,Efrati:2015eaa,Berthier:2015oma,Berthier:2015gja,Azatov:2015oxa,Englert:2015hrx,Buchalla:2015wfa,Buchalla:2015qju,Maltoni:2016yxb,Bjorn:2016zlr,Berthier:2016tkq,Englert:2017aqb,Brivio:2017bnu,Deutschmann:2017qum,deBlas:2017wmn,Alioli:2017ces,Murphy:2017omb,deBlas:2018tjm,Degrande:2018fog,Ellis:2018gqa,Hays:2018zze,Alves:2018nof,deBeurs:2018pvs,Hirschi:2018etq,Vryonidou:2018eyv,Bernlochner:2018opw}. A
priori, it is not clear whether new physics will first show up in the
behaviour of the Higgs boson rather than the top quark. Thus, a number
of more recent studies have looked at constraining effective theory in
the top quark
sector~\cite{Cao:2007ea,Degrande:2010kt,Zhang:2010dr,Greiner:2011tt,Degrande:2012wf,Jung:2014kxa,Dror:2015nkp,Davidson:2015zza,Rosello:2015sck,Bylund:2016phk}
(see
also~\cite{Grzadkowski:2003tf,Chen:2005vr,AguilarSaavedra:2008zc,AguilarSaavedra:2008gt,Nomura:2009tw,Hioki:2009hm,Hioki:2010zu,GonzalezSprinberg:2011kx,Fabbrichesi:2013bca,Hioki:2013hva,Aguilar-Saavedra:2014iga,AguilarSaavedra:2010nx,AguilarSaavedra:2011ct,Bernardo:2014vha,Fabbrichesi:2014wva,Cao:2015doa}
for analyses using the alternative language of anomalous couplings).

Ideally, one should attempt to constrain all possible higher
dimensional operators, using all possible experimental data. To this
end, Refs.~\cite{Buckley:2015nca,Buckley:2015lku} have presented a
proof of principle that such global fits are possible (see also
Refs.~\cite{Degrande:2014tta,Franzosi:2015osa,Zhang:2016omx,Degrande:2018fog,Mimasu:2015nqa,Degrande:2016dqg,Alioli:2018ljm,DHondt:2018cww,Rosello:2015sck,Bylund:2016phk}
for related work in both the top and Higgs sectors). In particular,
the fit of Ref.~\cite{Buckley:2015lku} directly constrained the
coefficients of all (combinations of) operators affecting single and
double top production and decay, as well as associated production of a
vector boson, using a wide variety of datasets from the Tevatron and
LHC (runs I and II). These datasets were all corrected back to parton
level, and compared with tree-level theory results, supplemented with
NLO information using (bin-by-bin) K-factors. There are many available
datasets, however, which do not admit such a simple theoretical
description. A typical example is observables sensitive to multiple
final state jets of QCD radiation, which cannot be accurately modelled
by available LO or NLO matrix elements. To this end, one may employ
parton shower algorithms to simulate additional radiation, and many
general purpose codes are available. Furthermore, parton showers may
be systematically matched to matrix elements calculated at
next-to-leading order in QCD perturbation
theory~\cite{Frixione:2002ik,Frixione:2003ei,Nagy:2005aa,Frixione:2007vw,Giele:2007di,Hoche:2010pf,Hoche:2010kg,Hoeche:2011fd},
and to algorithms which model the hadronisation of final state quarks
and gluons. Computations of this type for top-related processes,
including higher dimensional operators, have been recently presented
in Refs.~\cite{Bylund:2016phk,Maltoni:2016yxb}.

When many final state jets are required in a given observable, one
must systematically improve the output of a parton shower by including
higher order tree-level matrix elements. A matching prescription is
then needed to avoid double counting of radiation included in both the
matrix elements and the shower, and different schemes have been
presented in
Refs.~\cite{Catani:2001cc,Krauss:2002up,Lonnblad:2001iq,Lavesson:2005xu,Mangano:2001xp,Mangano:2002ea}. The
aim of such a matching procedure, roughly speaking, is to ensure that
jets that are widely separated from others in a given event are
generated by the hard matrix elements, and those which are
approximately collinear with other jets are generated by the parton
shower. The separation between these two regimes is specified by a
{\it matching scale} $Q$, and this must be carefully chosen so as to
avoid discontinuities in distributions relating to additional jet
radiation: too small, and matrix elements will be evaluated in
momentum regions in which they are becoming collinearly singular; too
large, and the parton shower will be used in kinematic regions where
it fails to approximate higher order matrix elements sufficiently
accurately. Such discontinuities thus represent a mismatch between how
QCD radiation is described by a parton shower, and by exact matrix
elements, and is a problem even within the SM alone.

When higher dimensional operators are added to the SM, a second source
of discontinuity arises. The operators generate additional Feynman
rules, including couplings to quarks and gluons. Thus, rates for the
emission of QCD radiation become modified. If matrix elements in this
theory are matched to a parton shower, it follows that widely
separated jets (generated by the matrix elements) will include the
effect of the BSM physics, whereas those which are not widely
separated will be generated by a parton shower which contains SM
radiation only, leading to a mismatch between the two descriptions.
Na\"{i}vely, one may expect this effect to be small: first of all, the
higher dimensional operators are typically suppressed by a large
energy scale. Moreover, they contain momentum-dependent numerators
that tend to boost radiated particles to higher transverse momenta,
thus widely separated from other particles. However, there is a more
formal argument one can give to explain why any additional
discontinuity resulting from the lack of EFT operators in the shower
is negligible relative to the SM effect, related to the fact that
emissions from EFT operators do not give rise to infrared
singularities. Although the ideas involved are well-known to a QCD
audience, this argument is not well-known in the BSM literature, and
thus we provide a detailed explanation in this paper.

Armed with results for matrix elements containing EFT corrections
matched to a parton shower one may address, for a particular
production process / observable requiring such a theory description,
whether or not useful additional constraints on given operators can be
obtained. Our motivation here is extending the global EFT fits of
Refs.~\cite{Buckley:2015nca,Buckley:2015lku} to include particle-level
observables, for which a notable example is top quark pair production
in association with multiple jets. For this process to provide a
worthwhile new input to a global EFT fit, it is important to check
that useful additional constraints are obtained by including radiation
observables as well as those related to the top particles
alone~\footnote{It is worth bearing in mind that there are good
  reasons for using a particle-level description even for inclusive
  top quark measurements, rather than parton level: such a comparison
  will be less sensitive to assumptions made in the unfolding step in
  the latter.}. We will study this in detail for a number of
operators, whose coefficients are set to values consistent with recent
constraints, finding that indeed there is more information to be
gleaned by adding observables sensitive to extra radiation.

The structure of the paper is as follows. In section~\ref{sec:theory},
we examine the role of effective theory corrections to the three-gluon
vertex, and argue that any additional discontinuity when matching
matrix elements and parton showers is kinematically subleading to the
existing SM discontinuity. In section~\ref{sec:results}, we study a
number of observables in top plus multijet production, and examine in
detail whether additional constraints can be obtained by using
observables sensitive to additional jet radiation, given the matching
uncertainties. Finally, we conclude in section~\ref{sec:conclude}.

\section{Matching effective theory matrix elements with parton showers}
\label{sec:theory}

\begin{figure}
\begin{center}
\scalebox{0.6}{\includegraphics{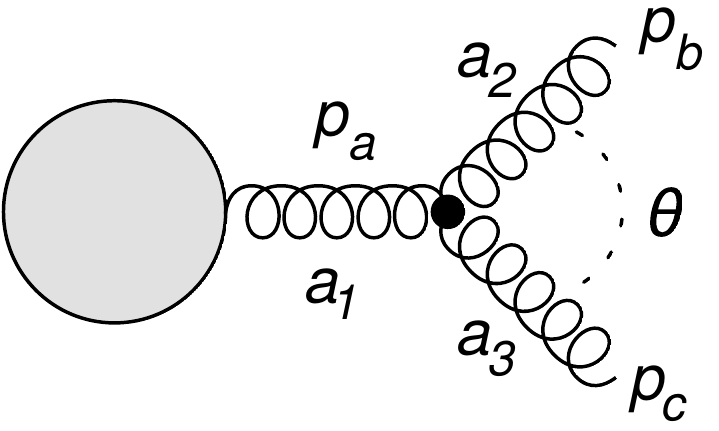}}
\caption{A gluon emerges from a scattering amplitude, before branching
  into two gluons, with angular separation $\theta$.}
\label{fig:gluonfig}
\end{center}
\end{figure}

As discussed above, if one matches higher order tree-level matrix
elements containing higher dimensional operators to a parton shower,
BSM effects are included in the matrix elements, but not the
shower. Information about the BSM physics is then missing in
additional jets that are generated by the shower, rather than the
matrix elements, where the former are formally correct only in the
collinear limit. This in turn means that jets which are sufficiently
collinear to other jets (defined in terms of the matching scale) in a
given event are potentially missing BSM effects. In this section, we
review in detail why this is not actually a problem in practice, due
to the differing nature of SM and BSM radiation in the collinear
limit.  Our presentation will be similar to that of
Refs.~\cite{Ellis:1991qj,Plehn:2009nd}. For illustrative purposes, we
consider the case of a gluon which branches into a gluon pair, with
momenta labelled as in Fig.~\ref{fig:gluonfig}.
The intermediate momentum $p_a=p_b+p_c$ has virtuality
\begin{equation}
t\equiv p_a^2\simeq z(1-z)E_a^2\theta^2,
\label{tdef}
\end{equation}
where
\begin{equation}
z=\frac{E_b}{E_a}=1-\frac{E_c}{E_a}
\label{zdef}
\end{equation}
is the fraction of the energy of parton $a$ that is carried by parton
$b$. Given that this virtuality is small as $\theta\rightarrow 0$, we
may treat parton $a$ as being approximately on-shell in the collinear
limit. We can then consider the amplitude for emission of a gluon from
the leg $p_a$, which will be given by
\begin{equation}
{\cal M}^{a_2,a_3,\ldots}_{n+1}(p_b,p_c,\ldots)=\epsilon^{\mu_1}(p_a)
\,\epsilon^{\mu_2}(p_b) \,\epsilon^{\mu_3}(p_c) V^{a_1 a_2 a_3}_{\mu_1
  \mu_2 \mu_3} (p_a,p_b,p_c){\cal M}^{a_1,\ldots}_n(p_a,\ldots),
\label{Mdef}
\end{equation}
where ${\cal M}_n$ is the amplitude before the gluon branching,
$\{a_i\}$ are adjoint indices of the gluons, and the ellipses denote
all momenta and colour degrees of freedom associated with partons that
are suppressed in Fig.~\ref{fig:gluonfig}. Furthermore,
$\epsilon^{\mu_i}$ is the polarisation vector of parton $i$, and
$V^{a_1 a_2 a_3}_{\mu_1 \mu_2 \mu_3}$ the three-gluon vertex. We may
write the latter as
\begin{equation}
V^{a_1a_2a_3}_{\mu_1 \mu_2 \mu_3}(p_1,p_2,p_3)=
V^{a_1a_2a_3}_{{\rm SM}, \mu_1 \mu_2 \mu_3}(p_1,p_2,p_3)
+V^{a_1a_2a_3}_{{\rm BSM}, \mu_1 \mu_2 \mu_3}(p_1,p_2,p_3),
\label{Vdef}
\end{equation}
where the first term on the right-hand side is the Standard Model
component (coming purely from the QCD Lagrangian), and the second term
collects the additional contributions arising from higher dimensional
operators. To this end let us consider the following three-gluon
operator:
\begin{equation}
{\cal O}_G=f_{a_1a_2a_3}\,G_\mu^{a_1\nu}\,G_\nu^{a_2\lambda}\,
G_{\lambda}^{a_3\mu},
\label{3Gop}
\end{equation}
where $G^A_{\mu\nu}$ is the gluon field strength tensor. We may also
consider the associated Lagrangian
\begin{equation}
{\cal L}_G=\frac{C_G}{\Lambda^2}\frac{1}{3!}{\cal O}_G,
\label{3GL}
\end{equation}
where $\Lambda$ is the new physics scale, and $C_G$ an unknown
coefficient. The effect of this Lagrangian is to generate new
interactions involving the gluon field, and in particular to modify
the Feynman rule for the three-gluon vertex. Explicit results for the
two terms on the right-hand side of Eq.~(\ref{Vdef}) are then 
\begin{equation}
V^{a_1a_2a_3}_{{\rm SM}, \mu_1\mu_2\mu_3}(p_1,p_2,p_3)=
g_s f_{a_1 a_2 a_3}\left[(p_1-p_2)_{\mu_3}\eta_{\mu_1\mu_2}
+(p_3-p_1)_{\mu_2}\eta_{\mu_1\mu_3}+(p_2-p_3)_{\mu_1}\eta_{\mu_2\mu_3}
\right]
\label{VSMdef}
\end{equation}
and
\begin{align}
&V^{a_1a_2a_3}_{{\rm BSM}, \mu_1\mu_2\mu_3}(p_1,p_2,p_3)=
\frac{C_G}{\Lambda^2}f^{a_1a_2a_3}\Big[
p_{1\mu_2}\,p_{2\mu_3}\,p_{3\mu_1}-p_{1\mu_3}\,p_{2\mu_1}\,p_{3\mu_2}
\notag\\
&\quad
+(p_1\cdot p_2)p_{3[\mu_2}\,\eta_{\mu_1]\mu_3}
+(p_1\cdot p_3)p_{2[\mu_1}\,\eta_{\mu_3]\mu_2}
+(p_2\cdot p_3)p_{1[\mu_3}\,\eta_{\mu_2]\mu_1}
\Big],
\label{VBSMdef}
\end{align}
where the square brackets in the subscripts denote antisymmetrisation
of indices. 

In constructing the squared amplitude in the SM, one need only include
diagrams in which the radiated gluon $p_b$ lands on the leg $p_a$ on
both sides of the final state cut. From Eq.~(\ref{Mdef}), the sum of
all such diagrams, summed over all final state polarisations and
averaged over the polarisation of the branching parton $p_a$, is (in
four spacetime dimensions)
\begin{equation}
\overline{|{\cal M}^{a_2,a_3,\ldots}_{n+1}(p_b,p_c,\ldots)|^2}=
\frac{1}{2}|{\cal M}^{a_1,\ldots}_n(p_a,\ldots)|^2\sum_{\rm pols}
\Big[\epsilon^{\mu_1}(p_a)
\,\epsilon^{\mu_2}(p_b) \,\epsilon^{\mu_3}(p_c) V^{a_1 a_2 a_3}_{\mu_1
  \mu_2 \mu_3} (p_a,p_b,p_c)\Big]^2.
\label{Mbar1}
\end{equation}
In order to evaluate this expression, one may choose an explicit basis
of polarisation vectors for each parton,
$\epsilon(p_i)\in\{\epsilon^{\rm in}_i,\epsilon^{\rm
    out}_i\}$, pointing in and out of the scattering plane
respectively. One may then use the dot
products~\cite{Ellis:1991qj,Plehn:2009nd}
\begin{align}
\epsilon^{\rm in}_i\cdot\epsilon^{\rm in}_j=
\epsilon^{\rm out}_i\cdot\epsilon^{\rm out}_j=-1,\quad
\epsilon^{\rm in}_i\cdot\epsilon^{\rm out}_j=
\epsilon^{\rm out}_i\cdot p_j=0,
\label{prods1}
\end{align}
as well as
\begin{align}
\epsilon_a^{\rm in}\cdot p_b\simeq -z(1-z)E_a\theta,\quad
\epsilon_b^{\rm in}\cdot p_c\simeq (1-z)E_a\theta,\quad
\epsilon_c^{\rm in}\cdot p_b\simeq -zE_a\theta 
\label{prods2}
\end{align}
and
\begin{align}
p_a\cdot p_b=p_a\cdot p_c=-p_b\cdot p_c\simeq 
-\frac{z(1-z)E_a^2\theta^2}{2},
\label{prods3}
\end{align}
to obtain
\begin{align}
\overline{|{\cal M}_{n+1}(p_b,p_c,\ldots)|^2}
&=\frac{4 C_A g_s^2}{t}\left[
\frac{1-z}{z}+\frac{z}{1-z}+z(1-z)\right.\notag\\
&\left.+\frac{C_G}{\Lambda^2}\frac{z(1-z)E_a^2\theta^2}{2}
-\left(\frac{C_G}{\Lambda^2}\right)^2
\frac{z^3(1-z)^3E_a^4\theta^4}{4}\right]
|{\cal M}^{a_1,\ldots}_n(p_a,\ldots)|^2.
\label{Mbar2}
\end{align}
The terms in the first line are recognisable as the usual QCD
splitting function $P_{gg}(z)$ describing the probability for a gluon
to branch into two gluons. The first term in the second line arises
from interference of the BSM contribution with the SM, and is
suppressed by the inverse square of the new physics scale. The second
term in the second line is quadratic in the new physics, and would mix
with potential dimension eight effects in the effective theory
expansion, thus is formally of higher order and can be neglected. It
is only the interference term that constitutes those BSM corrections
that are missing in the collinear region. However, upon studying this
term, we see explicitly that it contains a factor of $\theta^2$ and,
hence, is formally kinematically suppressed in the collinear
limit. From Eq.~(\ref{tdef}), the prefactor in Eq.~(\ref{Mbar2}) is
${\cal O}(\theta^{-2})$, and thus the SM term contains the well-known
collinear enhancement of QCD radiation. The BSM interference term
(including the prefactor) is ${\cal O}(\theta^0)$, and will be
negligible provided that the matching scale between matrix elements
and parton shower is chosen to be sufficiently small. Put another way,
any additional source of discontinuity in jet-related distributions
coming from the absence of BSM corrections in the shower, is
kinematically suppressed relative to the discontinuity already present
in the SM.

\begin{figure}
\begin{center}
\scalebox{1.0}{\includegraphics{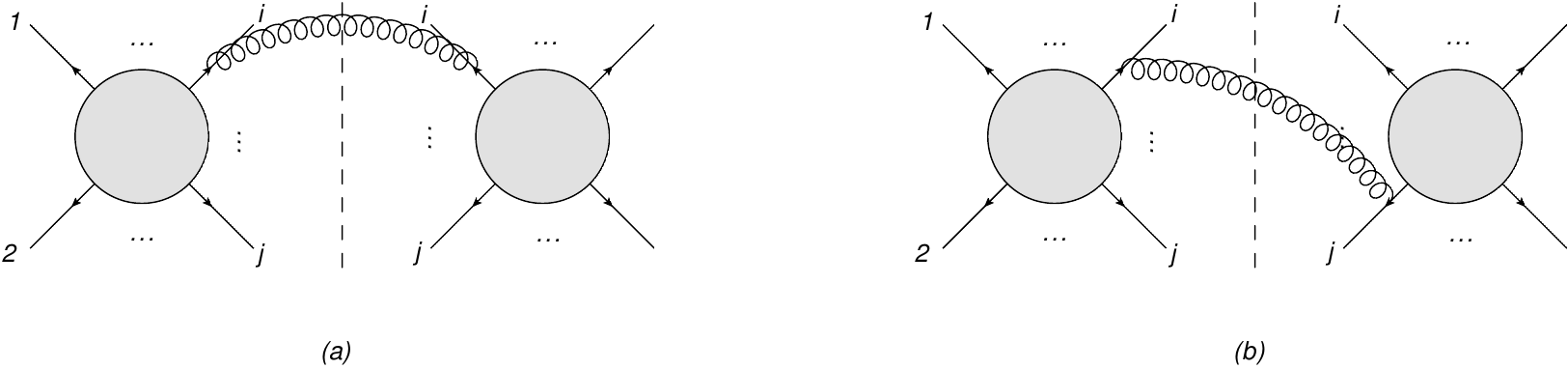}}
\caption{A gluon is emitted from parton leg $i$ and lands on leg $j$
  in the conjugate amplitude: (a) the case $i=j$; (b) the case $i\neq
  j$.}
\label{fig:diags}
\end{center}
\end{figure}

Given the above discussion, one may ponder whether it is possible to
nevertheless include the BSM interference contribution in
Eq.~(\ref{Mbar2}) in the gluon branching probability, despite the fact
that this corresponds (eventually) to resumming a subleading
contribution. However, such a procedure would be formally
incorrect. In general, the radiated gluon may be emitted from parton
leg $i$, and land on leg $j$ in the conjugate amplitude. Above, we
have considered only contributions for which $i=j$, as shown in
Fig.~\ref{fig:diags}(a). 
This is correct for the SM: such contributions, as we have seen above,
are ${\cal O}(\theta^{-2})$, whereas diagrams with $i=j$
(Fig.~\ref{fig:diags}(b)) are ${\cal O}(\theta^0)$. It is
incomplete, however, for the BSM interference contributions. Diagrams
with $i=j$ and $i\neq j$ are both of the same kinematic order (${\cal
  O}(\theta^0)$). All of them must be therefore be included to ensure a
gauge invariant result, so that it makes no sense to resum only a
subset of them.

Above we have examined only the operator of Eq.~(\ref{3Gop}). However,
the kinematic suppression that we have observed will be fully general,
including operators that affect other parton branchings, involving
(anti-)quarks in addition to gluons. This follows from the fact that
any dimension six operator appears in the Lagrangian with a
dimensionless coefficient $C_i$, and an inverse power of the new
physics scale, $\Lambda^{-2}$. For the Lagrangian to have the correct
dimension, the momentum space Lagrangian must contain two powers of
momentum, that combine with $\Lambda$ to make a dimensionless
ratio. We can see this explicitly in the example of the three-gluon
operator given above: the BSM vertex of Eq.~(\ref{VBSMdef}) contains
two additional powers of momentum relative to the SM result of
Eq.~(\ref{VSMdef}). When evaluating the graph of
Fig.~\ref{fig:gluonfig}, there is only one momentum scale available,
namely the virtuality of the branching parton, $p_a^2$. Thus, the
higher dimensional operator must contribute an interference term
\begin{equation}
\sim\frac{p_a^2}{\Lambda^2}\sim \frac{z(1-z)E_a^2\theta^2}{\Lambda^2},
\label{dimscale}
\end{equation}
where we have used Eq.~(\ref{tdef}). This is indeed observed in
Eq.~(\ref{Mbar2}). 

We have so far seen that the effects of higher dimensional operators
are negligible in the collinear region, and thus should not
significantly contribute to discontinuities in jet-related kinematic
distributions when matching parton showers (based on enhanced
collinear radiation) to matrix elements. However, this is not the full
story. Parton shower algorithms also include the effects of wide-angle
soft radiation, by e.g. explicit angular ordering, or the choice of
evolution variable~\cite{Ellis:1991qj,Plehn:2009nd}. The above
dimensional argument applies also in this case: for soft, but not
necessarily collinear, emissions, the only momentum scale that can
combine with the new physics scale in the BSM vertex for the radiated
parton is the virtuality of the emitting parton. This remains small if
the emitted parton is soft. Indeed, looking at the interference term
in Eq.~(\ref{Mbar2}), we see that this vanishes as $z\rightarrow 0$ or
$z\rightarrow 1$, corresponding to the two limits in which either
parton $a$ or parton $b$ is soft. There is thus no soft singularity
from the BSM part, mirroring the lack of collinear
singularity. {Furthermore, the fact that the virtuality of the
  emitting parton is the only relevant scale that can combine with the
  new physics scale in the soft or collinear region means that the
  above discussion is fully general, and applies to all QCD dimension
  six operators.}

In this section, we have reviewed in detail the fact that additional
radiation produced by EFT operators is not associated with collinear
singularities, and thus does not lead to SM-like discontinuities when
matching tree-level matrix elements with a parton shower. However, in
order to obtain meaningful constraints from observables requiring such
a theory description, it is important to examine whether or not
deviations due to EFT corrections lead to significant new information,
in light of the theoretical uncertainties due to the matching
procedure. This is the subject of the following section.

\section{Results}
\label{sec:results}

As discussed above, the aim of our study is to ascertain whether or
not jet observables in top production provide additional constraints
on new physics (as described using EFT), relative to observables only
involving the top quarks 
(for similar analyses of BSM QCD multi-jet production see~\cite{Krauss:2016ely}).
To this end, we must first examine the
matching uncertainty affecting how the jets are modelled. 

\subsection{Effect of dimension six operators on jet radiation}
\label{sec:EFTradiation}

In this section, we present a number of example distributions,
including EFT effects consistent with current constraints on operator
coefficients~\cite{Buckley:2015lku}. Results are obtained as
follows. We implement EFT operators in a
\texttt{FeynRules}~\cite{Alloul:2013bka} model file, and interface
this with \texttt{MadGraph5\_aMC@NLO}~\cite{Alwall:2014hca} for the
generation of tree-level events containing top pairs with up to 2
jets. These are matched to the parton shower \texttt{Pythia
  8}~\cite{Sjostrand:2007gs}, using the default MadGraph MLM-based
matching scheme~\cite{Alwall:2007fs}, with a central matching scale
$Q=30$ GeV. Our default choice for the renormalisation and
factorisation scales is the top mass, $\mu_{\rm ren}=\mu_{\rm
  fact.}=m_t$, and we use the parton distributions of
Ref.~\cite{Ball:2012cx}. We cluster all visible final-state particles
into jets using the anti-$k_T$ algorithm~\cite{Cacciari:2008gp} with
jet radius $R=0.4$, as implemented in
\texttt{FastJet}~\cite{Cacciari:2011ma}. We consider only the dilepton
final state, and require both leptons to be isolated from hadronic
activity, defined via the requirement that the total transverse
momentum with a cone of radius $\Delta R=0.3$ around the lepton
satisfies $p_T^{\rm cone}\leq 0.1 \times p_{T,l}$, where $p_{T,l}$ is the
transverse momentum of the lepton. We remove the isolated leptons and
any $b$-tagged jets from the list of final state jets and particles,
so as to consider only jets originating from additional radiation.

\begin{figure}
\centering
\begin{subfigure}{.5\textwidth}
  \centering
  \scalebox{0.85}{\includegraphics{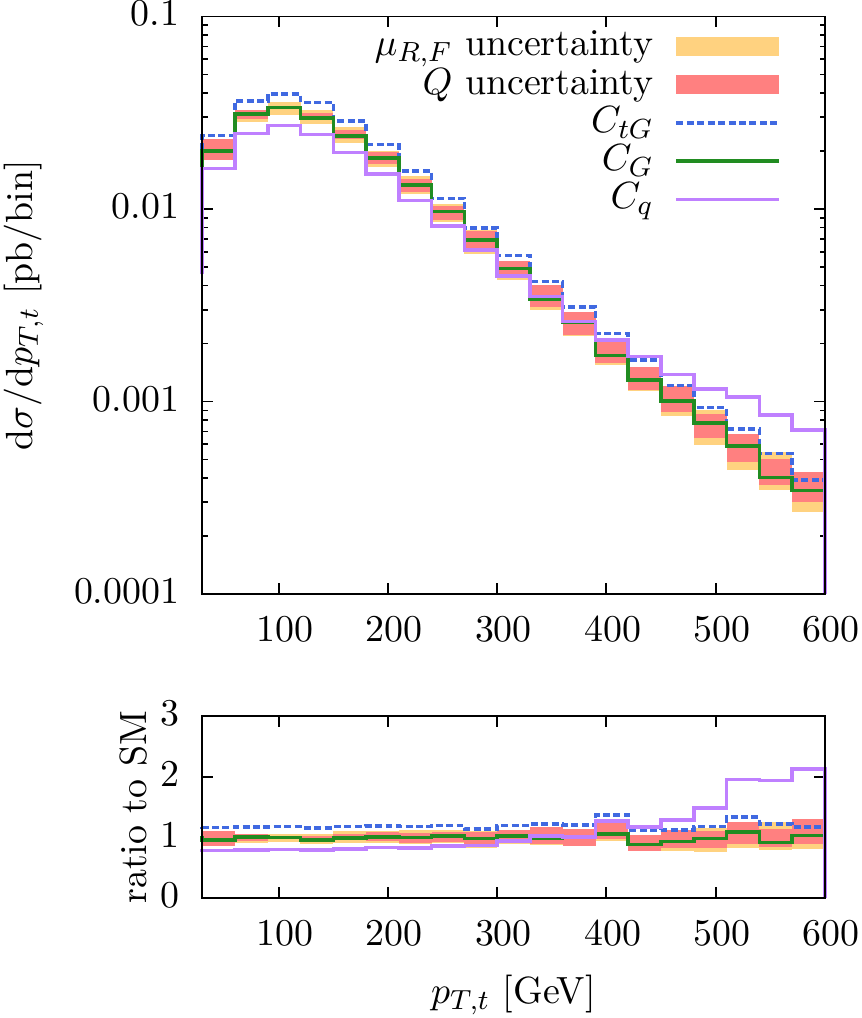}}
  \caption{}
\end{subfigure}%
\begin{subfigure}{.5\textwidth}
  \centering
  \scalebox{0.85}{\includegraphics{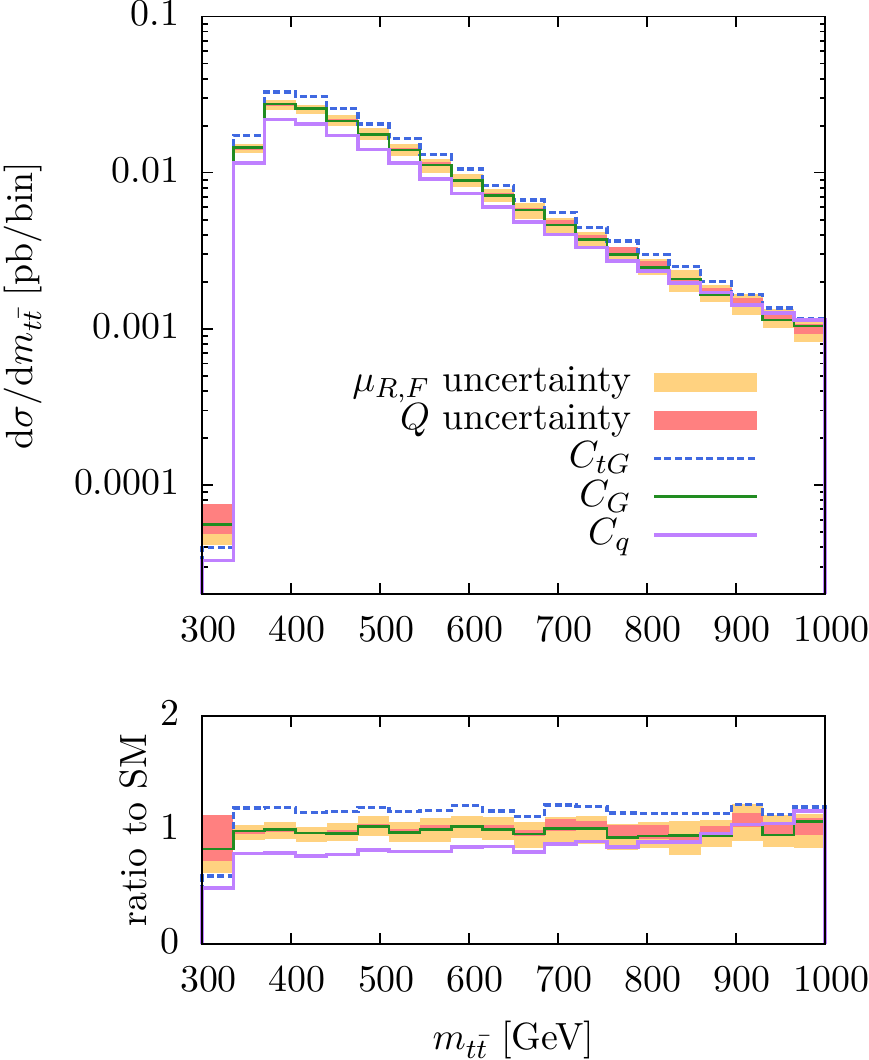}}
  \caption{}
\end{subfigure}
\caption{(a) The transverse momentum of the hardest top particle, for
  both the SM only, and including separately the operators of
  Eqs.~(\ref{Cq})--(\ref{CuG}), with coefficients set as in
  Eq.~(\ref{constraints}); (b) similar results for the top pair
  invariant mass.}
\label{fig:mtt}
\end{figure}

There are six combinations of dimension 6 EFT operators affecting top
quark pair production at tree-level in the SMEFT (see
e.g.~\cite{AguilarSaavedra:2018nen}), for which we use the Warsaw
basis of Ref.~\cite{Grzadkowski:2010es}. Four of these are 4-fermion
operators, for which we we choose the representative example
\begin{equation}
{\cal O}_q=\frac{C_q}{\Lambda^2}\left(\bar{u}\gamma^\mu\, u
+\bar{d}\gamma^\mu d\right)\left(\bar{t}\gamma_\mu t\right)
\label{Cq}
\end{equation}
in what follows (n.b. this consists of a sum of operators appearing in
Ref.~\cite{Grzadkowski:2010es}). Here $u$, $d$ and $t$ are the up,
down and top quark fields respectively. The remaining two operators
are a correction to the three-gluon vertex~\footnote{The interference
  of the top pair production amplitude containing the gluon operator
  of Eq.~(\ref{CG}) with the corresponding SM amplitude vanishes, such
  that this operator contributes at quadratic, and thus dimension
  eight, level only. This leads some people to disregard this
  operator, but a different school of thought is that it should be
  included as the leading contribution of this operator to the given
  process. We choose to follow the latter approach.}
\begin{equation}
{\cal O}_G=\frac{C_G}{\Lambda^2}f_{ABC}\,G_\mu^{A\nu}\,G_\nu^{B\lambda}
\,G_\lambda^{C\mu}
\label{CG}
\end{equation}
(where $G^A_{\mu\nu}$ is the gluon field strength tensor and $f_{ABC}$
the structure constants of the SU(3) gauge group), and the
chromomagnetic moment operator
\begin{equation}
{\cal O}_{uG}=\frac{C_{uG}}{\Lambda^2}(\bar{q}\sigma^{\mu\nu} T^A u)\tilde{\varphi}\,
G^A_{\mu\nu},
\label{CuG}
\end{equation}
where $\sigma^{\mu\nu}$ is a fermionic spin generator, and $T^A$ an
SU(3) generator. The combinations
\begin{equation}
\tilde{C}_i=\frac{C_i}{\Lambda^2}
\end{equation}
have already been significantly constrained by global analyses of top
quark data. Motivated by the analysis of Ref.~\cite{Buckley:2015lku},
we take as representative constraints
\begin{equation}
\tilde{C}_{q}=1.25\, {\rm TeV}^{-2},\quad
\tilde{C}_{G}=0.45\, {\rm TeV}^{-2},\quad 
\tilde{C}_{tG}\equiv \tilde{C}_{uG}^{33}=0.64\, {\rm TeV}^{-2}.  
\label{constraints}
\end{equation}
In Fig.~\ref{fig:mtt}, we show the distribution of the transverse
momentum $p_{T,t}$ of the hardest top particle~\footnote{{We
    extract the top particles from the Monte Carlo event record, such
    that they correspond to the individual top jets before showering
    and decay.}}, together with the invariant mass $m_{t\bar{t}}$ of
the top quark pair. Note that the former differs from the top
transverse momentum used in the fit of Ref.~\cite{Buckley:2015lku},
which used data corrected back to parton level, where extra radiation
had been accounted for in the unfolding process. For such datasets
(which mimic the $2\rightarrow2$ scattering process), the transverse
momentum distributions of the top and antitop quarks will be
equal. When extra radiation is involved, the symmetry between the top
and antitop $p_T$ distributions is broken, and one may choose whether
to isolate the transverse momentum of the top quark (rather than
antitop), or to take the hardest top particle. A reason to use the
latter is that it should be more sensitive to details of the
additional radiation, given that it amplifies the recoil of the top
(or antitop) against the extra jets. By contrast, the invariant mass
distribution is more stable against radiative corrections.

In each panel of Fig.~\ref{fig:mtt}, the orange and red bands depict
the (renormalisation and factorisation) scale and matching scale
uncertainties associated with the SM result. We see in both cases that
the matching uncertainty is smaller than the other scale variation,
suggesting that modelling of additional radiation is well under
control. Both the gluon and four-fermion operators show a shape
difference with respect to the SM, albeit slight in the former case
given that the gluon operator is already rather
well-constrained. Nevertheless, it is difficult to gauge the
statistical significance of the deviation from the SM by eye alone,
and a much more quantitative description will be provided in the
following section. The effect of the four-fermion operator in the
$p_{T,t}$ spectrum is sizeable at large transverse momenta, as
expected given that EFT operators often boost final state particles,
as they contain extra momenta to offset the factor inverse new physics
scale $\Lambda^{-2}$. {For four-fermion operators, one may
  understand the enhancement as effectively arising from the lack of a
  SM propagator factor $1/q^2$.} Furthermore, the transverse momentum
of the hardest top particle should be particularly sensitive to the
nature of additional radiation, as discussed above. Deviations are
less evident, as expected, in the invariant mass spectrum, although
there is still a deviation from the SM, which is worth quantifying
further. Note that the dipole operator has a very similar shape to the
SM contribution (as noted in Ref.~\cite{Degrande:2010kt}), but
nevertheless leads to a change in overall normalisation that is still
compatible with current constraints.

\begin{figure}
\centering
\begin{subfigure}{.5\textwidth}
  \centering
  \scalebox{0.85}{\includegraphics{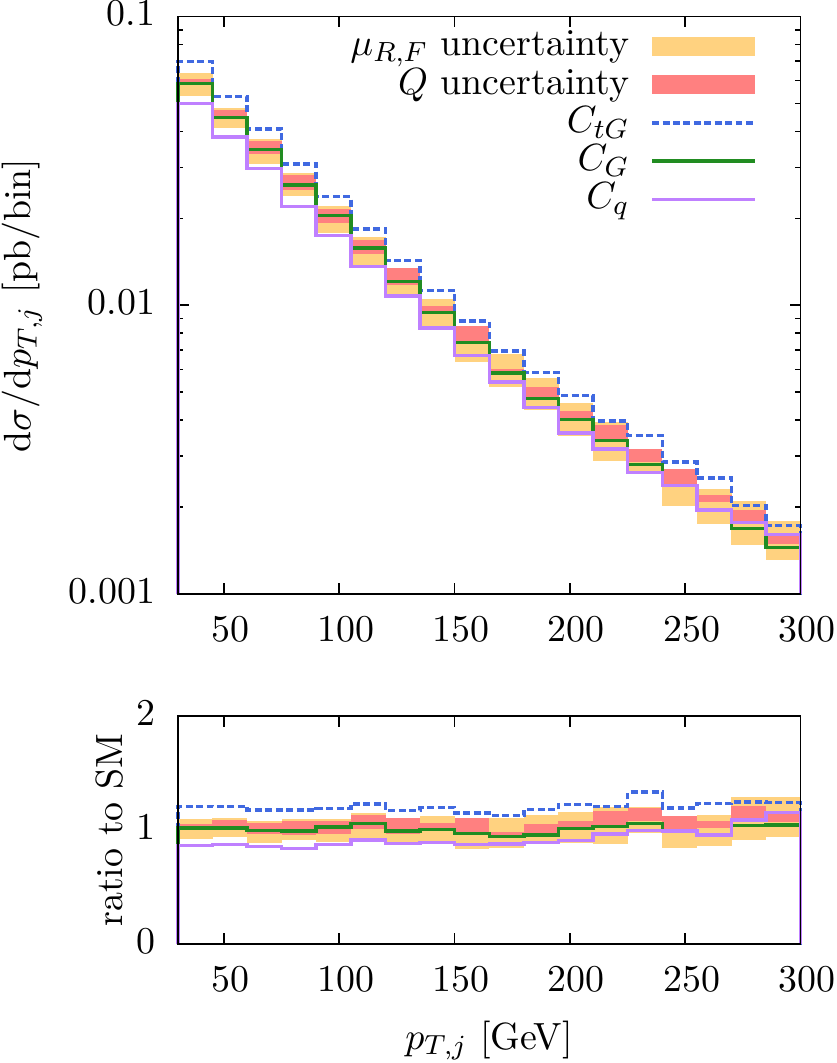}}
  \caption{}
\end{subfigure}%
\begin{subfigure}{.5\textwidth}
  \centering
  \scalebox{0.85}{\includegraphics{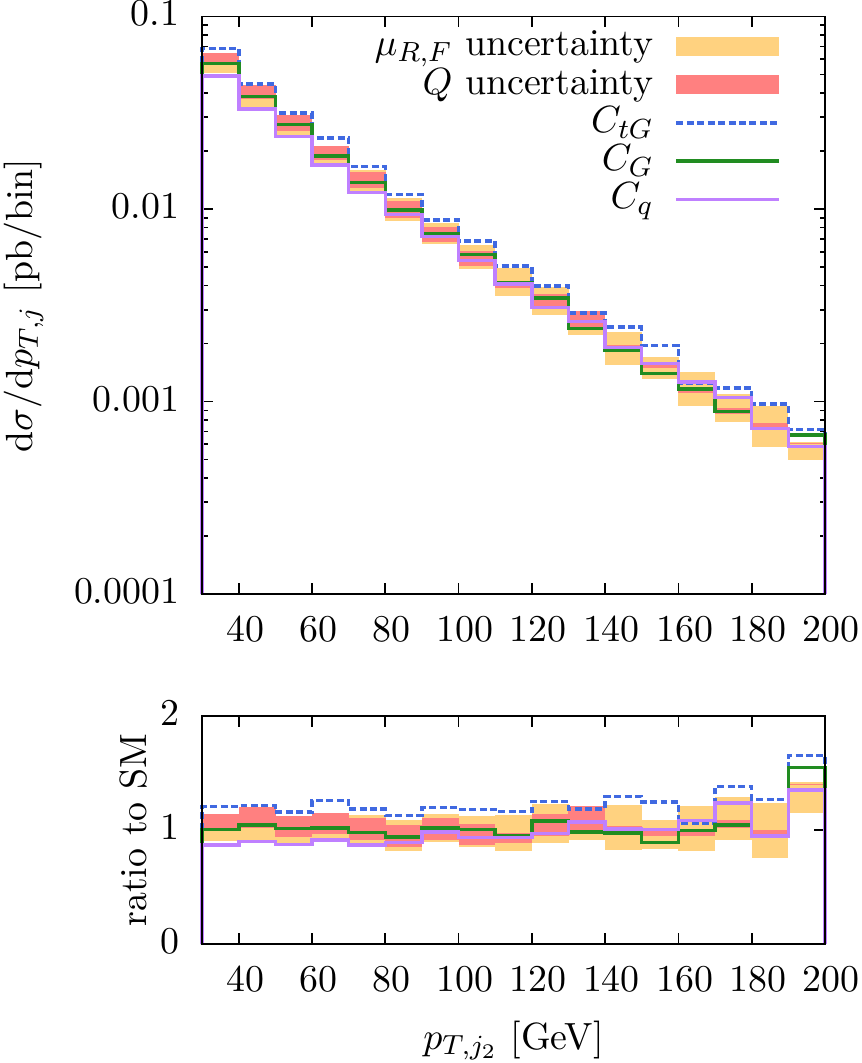}}
  \caption{}
\end{subfigure}
\caption{(a) The transverse momentum of the first hardest additional
  jet, for both the SM only, and including separately the operators of
  Eqs.~(\ref{Cq})--(\ref{CuG}), with coefficients set as in
  Eq.~(\ref{constraints}); (b) similar results for the second hardest
  additional jet.}
\label{fig:ptjets}
\end{figure}

In Fig.~\ref{fig:ptjets} we show the transverse momenta of the first
and second hardest additional jets, using the same conventions as
Fig.~\ref{fig:mtt}. Again we see that the matching uncertainty is
smaller than the other scale uncertainties, and that are potentially
statistically significant deviations from the SM.
It is interesting, however, to note that the effect of the four
fermion operator is much smaller than for the transverse momentum
spectrum of the hardest top particle at high $p_T$. This can be at
least partly explained from the fact that top pair production at the
LHC is dominated by the gluon channel. Thus, when only the
four-fermion operator is switched on, most of the additional radiation
will be purely SM-like. Another feature of Fig.~\ref{fig:ptjets} is
that the dipole operator of Eq.~(\ref{CuG}) leads to a normalisation
change of the jet radiation profile, but not a significant shape
change. It thus mirrors the properties already observed for
top-related observables in Ref.~\cite{Degrande:2010kt}, that the shape
of kinematic distributions involving the dipole operator is highly
similar to the SM alone. Thus, we see that the transverse momenta of
the additional jets are in principle useful for distinguishing the
dipole and three-gluon operators, whilst providing complementary
information to those observables (such as $m_{t\bar{t}}$) that also
constrain four fermion operators.

\begin{figure}
\centering
\begin{subfigure}{.5\textwidth}
  \centering
  \scalebox{0.85}{\includegraphics{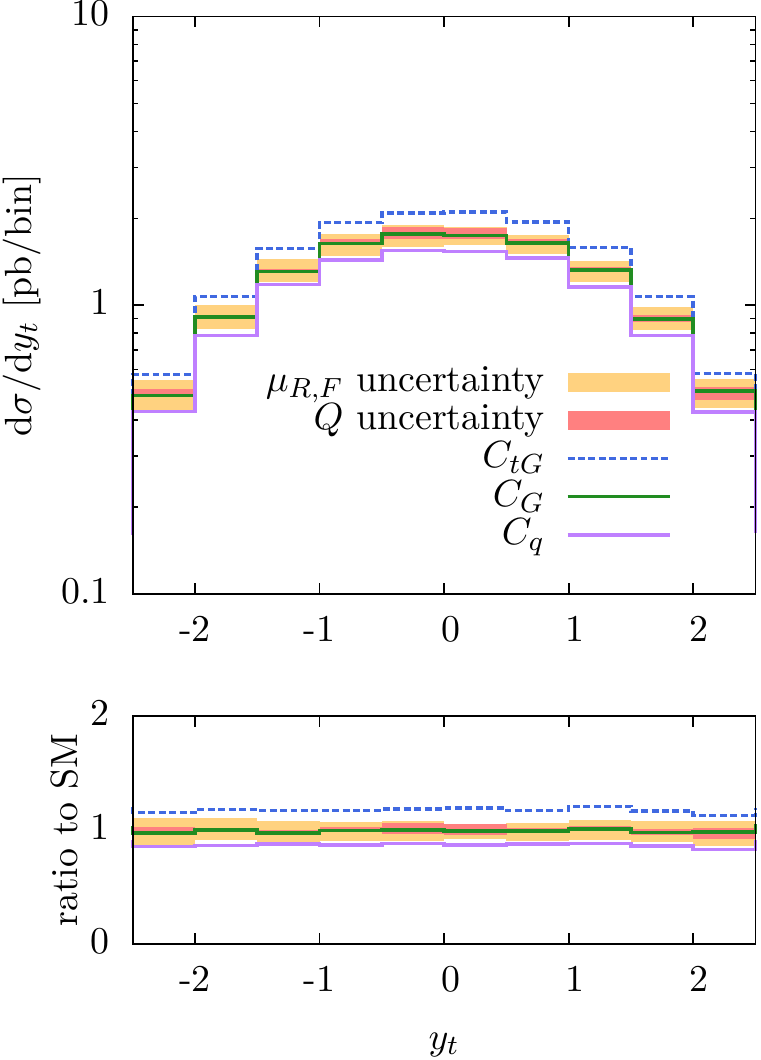}}
  \caption{}
\end{subfigure}%
\begin{subfigure}{.5\textwidth}
  \centering
  \scalebox{0.85}{\includegraphics{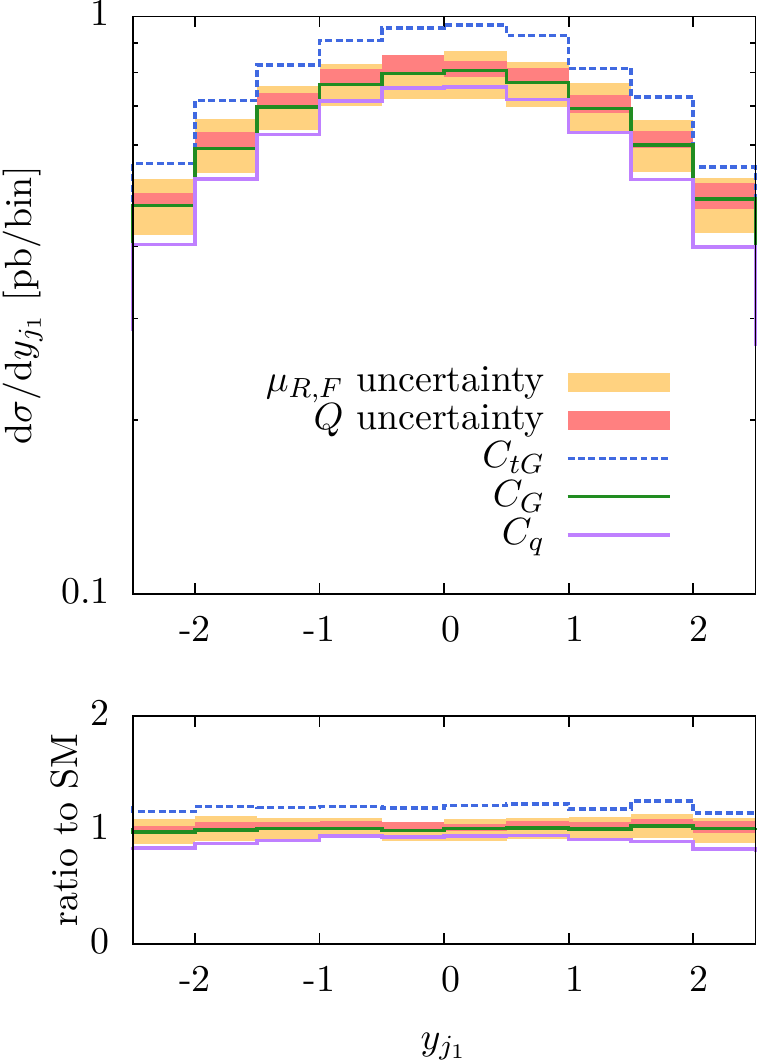}}
  \caption{}
\end{subfigure}
\caption{(a) The rapidity of the top quark, for both the SM only, and
  including separately the operators of Eqs.~(\ref{Cq})--(\ref{CuG}),
  with coefficients set as in Eq.~(\ref{constraints}); (b) similar
  results for the hardest additional jet.}
\label{fig:rapidity}
\end{figure}

In Fig.~\ref{fig:rapidity}, we show the rapidity of the top quark,
and of the hardest additional jet
(similar results are obtained for the second hardest jet).
The results are consistent with previous plots: the effect of the
dipole operator is to change the normalisation of the SM contribution,
but not the shape. The four fermion operator again has a smaller
effect, although becomes marginally more pronounced at higher absolute
rapidities, due to the fact that the parton luminosity then diminishes
the gluon initiated channel.

To summarise, we have seen that EFT contributions to observables
sensitive to additional radiation in top pair production indeed lead
to deviations from the SM, and of comparable size to those observables
(e.g. top transverse momentum, and the top pair invariant mass) that
are already used in EFT fits in the top sector. Furthermore, these
deviations survive against the matching and scale uncertainties
associated with the SM results. How much discriminating power these
extra observables have depends on the amount of data collected in
coming years, but also on whether the additional jet observables are
highly correlated with the top quark kinematics. We explore these
issues in the following section.

\subsection{Distinguishing power of jet observables}
\label{sec:distinguish}

Above, we have seen that EFT operators significantly affect additional
jet radiation in top pair production, such that observables involving
this radiation can potentially provide useful additional constraints
in global fits of top quark EFT to data. In order to check whether or
not this is realised, however, we must examine the degree of
correlation between observables involving the jet radiation, and those
involving the top particles alone. There is clearly {\it some} degree
of correlation, given that the top and antitop will recoil against
additional radiation. However, it may well be the case that certain
top observables are less correlated with radiation properties than
others. This is then useful information for choosing optimal (i.e. the
most complementary) sets of observables with which to constrain new
physics.

\begin{figure}
\centering
\begin{subfigure}{.5\textwidth}
  \centering
  \scalebox{0.85}{\includegraphics{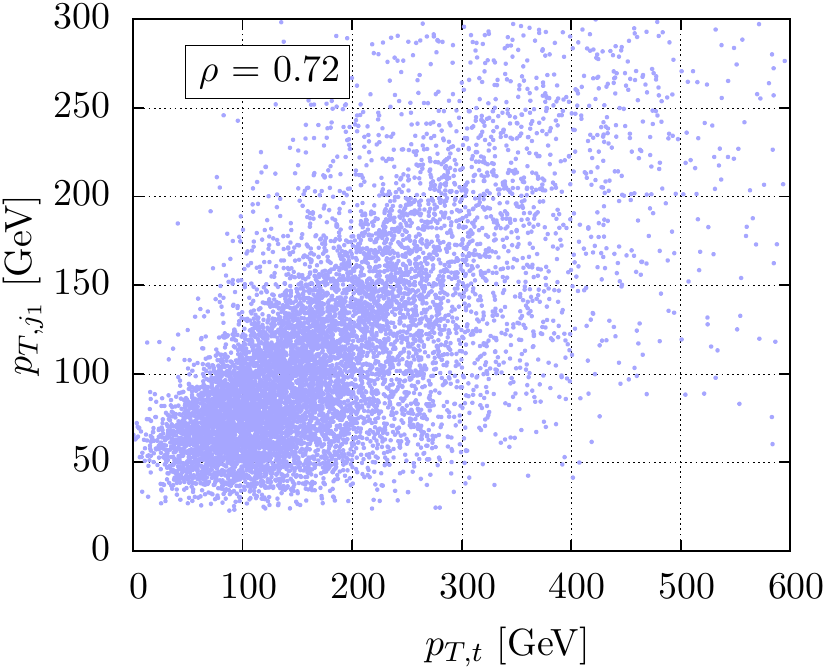}}
  \caption{}
\end{subfigure}%
\begin{subfigure}{.5\textwidth}
  \centering
  \scalebox{0.85}{\includegraphics{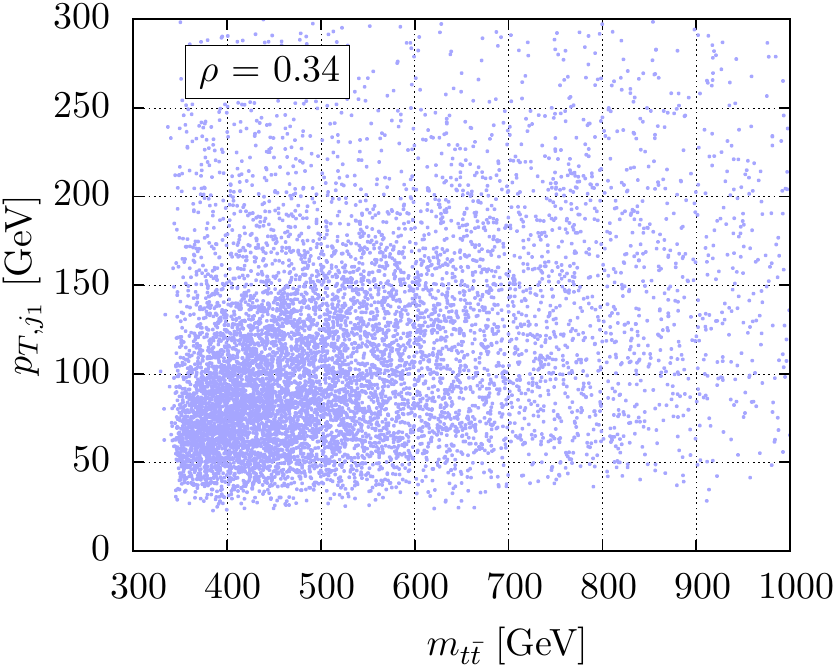}}
  \caption{}
\end{subfigure}
\caption{Scatter plots showing the transverse momentum of the hardest
  additional jet vs. (a) the transverse momentum of the hardest top
  particle; (b) the top pair invariant mass.}
\label{fig:jet1corr}
\end{figure}

\begin{figure}
\centering
\begin{subfigure}{.5\textwidth}
  \centering
  \scalebox{0.85}{\includegraphics{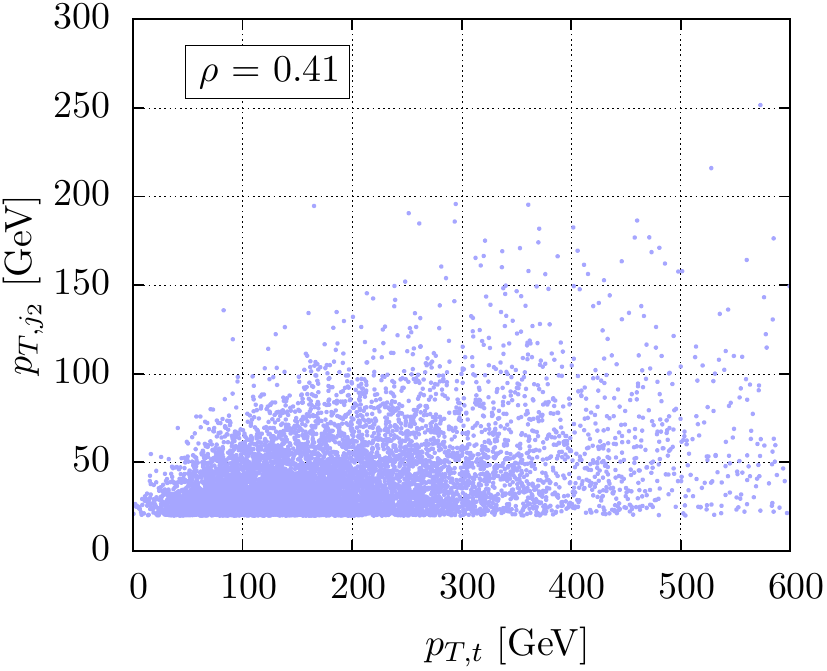}}
  \caption{}
\end{subfigure}%
\begin{subfigure}{.5\textwidth}
  \centering
  \scalebox{0.85}{\includegraphics{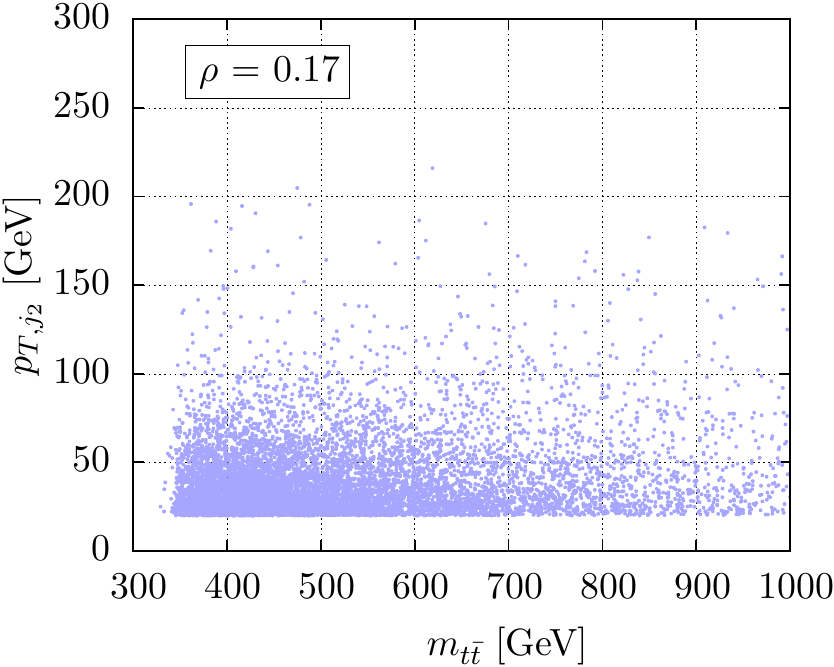}}
  \caption{}
\end{subfigure}
\caption{Scatter plots showing the transverse momentum of the second
  hardest additional jet vs. (a) the transverse momentum of the
  hardest top particle; (b) the top pair invariant mass.}
\label{fig:jet2corr}
\end{figure}

In Fig.~\ref{fig:jet1corr}, we show two-dimensional scatter plots of
the $p_T$ of the first hardest jet, and either the $p_T$ of the
hardest top particle, or the top pair invariant mass. All results are
calculated in the SM only.
In each plot, we also show the Pearson correlation coefficient
$\rho$. We see that the transverse momentum of the hardest top
particle is more correlated with the properties of the hardest jet
than the top pair invariant mass is, as expected given that the latter
observable is more stable to higher order corrections. In the case of
the invariant mass, the correlation coefficient is less than 0.5,
suggesting that indeed the additional jet radiation is capable of
providing significant complementary information relative to top
properties alone. Similar plots are shown in Fig.~\ref{fig:jet2corr}
for the second hardest jet.
Unsurprisingly, the top properties are less sensitive to the second
hardest jet than they are to the first hardest jet. Again, we see that
the invariant mass provides the most complementary information to the
the properties of the jet radiation.

Let us now turn to the question of how sensitivity to new physics is
affected by the inclusion of jet radiation observables. To estimate
the gain in sensitivity, we perform a binned hypothesis test using
two-dimensional distributions based on pairs of observables discussed
above (in principle three-dimensional distributions carry even more
shape information, but suffer from poor statistics). Our hypothesis
test is based on the modified frequentist $CL_s$
method~\cite{Junk:1999kv}. We take the signal hypothesis ($s$) to be
each dimension six operator in turn. The background hypothesis ($b$)
is the SM only, and we generate (pseudo)-data corresponding to the
background, before calculating the binned log-likelihood ratio
\begin{equation}
q \equiv -2\log Q = -2 \sum_{\substack{i}} \left(-s_i + d_i \log\left(1+\frac{s_i}{b_i}\right) \right),
\label{qdef}
\end{equation}
where the sum is over bins $i$, $s_i$ and $b_i$ are the expected
number of signal and background values respectively, and $d_i$ is the
number of observed events. The confidence levels for excluding the
$s+b$ and $b$-only hypotheses are
\begin{equation}
CL_{s+b} = P_{s+b}(q > q_{obs}), \qquad CL_{b} = P_{b}(q > q_{obs}).
\end{equation}
These represent, respectively, the probability that the test statistic
$q$ would be greater than that observed in the data, given the
hypothesised number of signal and background events $s+b$ or
background-only events $b$. In practice, we numerically evaluate these
$p$-values by generating a large number of Monte Carlo
pseudo-experiments, with $CL_{s+b}$ being the fraction of
pseudo-experiments that generate \emph{at least as many} events as
observed in the data. A signal hypothesis is regarded as excluded at
the 95\% confidence level if $CL_s\equiv CL_{s+b}/(1-CL_b) \le 0.05$.

To judge the usefulness of observables sensitive to additional jet
radiation, we take observables $X\in\{m_{t\bar{t}}, p_{T,t}\}$, and
calculate the $CL_s$ in two cases: (i) using $X$ alone; (ii) using $X$
in combination with the transverse momentum of the first hardest jet,
$p_{T,j_1}$. We then examine the ratio
\begin{equation}
\frac{CL_s(X,p_{T,j_1})}{CL_s(X)},
\label{improvement}
\end{equation}
which measures the ``improvement'' due to including the extra
radiation. We choose $p_{T,j_1}$ as a particular example, but similar
results are obtained by choosing other radiation observables. Given
that the dipole and gluon operators of Eqs.~(\ref{CG}), (\ref{CuG})
appear, from Figs.~\ref{fig:mtt}--\ref{fig:rapidity}, to be the
hardest to distinguish from the SM, we will focus on these.

\begin{figure}
\centering
\scalebox{0.85}{\includegraphics{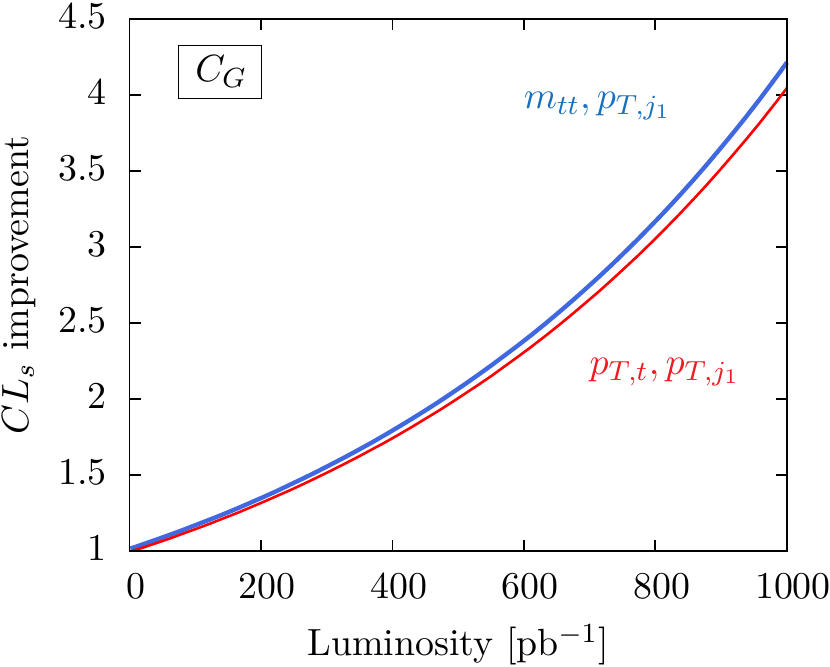}}
\hspace{1cm}
\scalebox{0.85}{\includegraphics{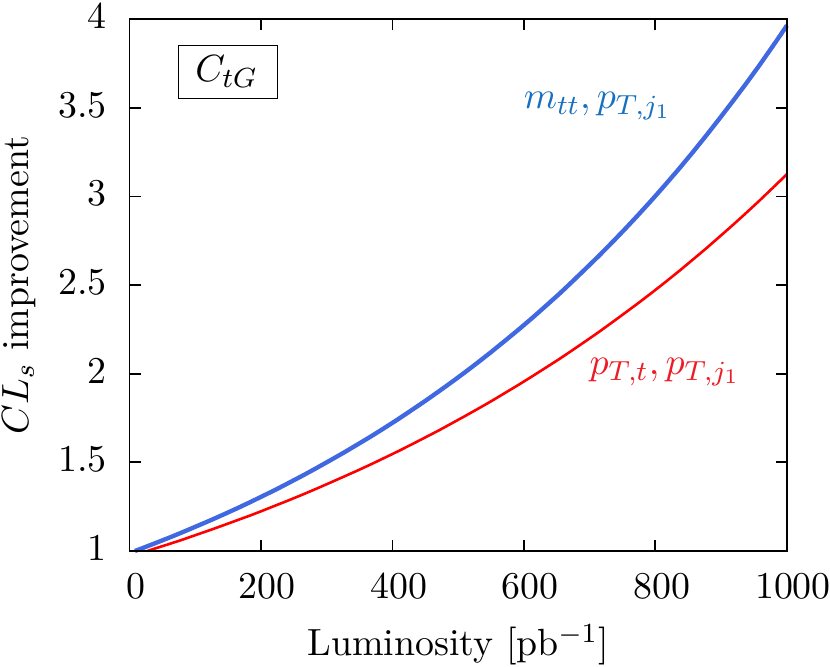}}
\caption{Relative improvement in the $CL_s$ for the EFT signal plus SM
  background, for the dipole and gluon operators of Eqs.~(\ref{CG}), 
  (\ref{CuG}), when using the transverse momentum of the hardest
  radiated jet in additional to: the top pair invariant mass (blue
  line); the transverse momentum of the hardest top particle (red
  line).}
\label{fig:clsplots}
\end{figure}

Results for each operator are shown in Fig.~\ref{fig:clsplots},
where to obtain the luminosity scale on the horizontal axis, we have
multiplied all cross-sections by a factor of $6\%$, corresponding to a
typical event selection efficiency for dileptonic top pair events. 
We see in both cases that using the additional jet radiation leads to
a significant improvement in the $CL_s$. More improvement is obtained
when adding the radiation to the invariant mass distribution rather
than the $p_T$ of the hardest top, as expected given that the former
is less correlated with the radiation. More improvement is seen for
the gluon operator, presumably due to the fact that this leads to a
significant shape change with respect to the SM, whereas the dipole
operator does not.

We did not present results for the four fermion operator of
Eq.~(\ref{Cq}). The negative interference in some kinematic regions
means that distributions involving this operator sometimes undershoot,
and sometimes overshoot the SM, as can be clearly seen in
Figs.~\ref{fig:mtt}--\ref{fig:rapidity}. This in turn leads to
cancellations in the log likelihood ratio of Eq.~(\ref{qdef}), so that
this is not the best quantity to use to measure the advantage of using
additional information. One could instead use e.g. $\chi^2$ values,
although it is any case already clear from
Figs.~\ref{fig:mtt}--\ref{fig:rapidity} that the four fermion
operator typically leads to much larger deviations from the SM subject
to current constraints, thereby rendering the analysis of this section
less relevant.

\section{Conclusion}
\label{sec:conclude}

In this paper, we have considered the issue of whether observables
relating to additional jet radiation in top pair production provide a
useful input to global fits of effective field theory (EFT) in the top
sector. In the absence of NLO QCD corrections to processes containing
EFT operators, the best way of describing such observables is to use
higher order tree-level matrix elements interfaced with a parton
shower. One may then worry about potential discontinuities arising
from the fact that radiation generated by the matrix elements includes
BSM effects, whereas radiation generated by the shower does not. We
have reviewed in section~\ref{sec:theory} why this is not a problem in
practice, due to the fact that the new physics contributions do not
generate soft or collinear singularities.

We then studied top pair production generated at tree-level with up to
two additional jets, matched to a parton shower. The matching
uncertainty was found to be smaller than the factorisation and
renormalisation scale uncertainty. Furthermore, deviations from the SM
due to EFT operators could be observed in a number of kinematic
distributions, including those associated with additional jet
radiation. We quantified this in section~\ref{sec:distinguish} by
looking at the relative improvement in the $CL_s$ for the EFT signal
plus SM background, when using the $p_T$ of the hardest additional jet
in addition to the top pair invariant mass or $p_T$ of the hardest top
particle. We saw significant improvements, suggesting that indeed
multijet observables can provide highly useful complementary
information to inclusive top observables alone. The inclusion of such
observables in global EFT fits is in progress.


\section*{Acknowledgments}
We are very grateful to Andy Buckley for detailed comments and help
regarding statistics issues. Furthermore, we thank Jay Howarth for
input regarding the event selection efficiency for dileptonic top pair
events, and Keith Hamilton for further useful discussions.  CDW is
supported by the UK Science and Technology Facilities Council (STFC),
under grant ST/P000754/1. CE is supported by the IPPP Associateship
scheme, and by the STFC under grant ST/P000746/1.


\bibliography{paper}

\end{document}